\documentstyle[12pt]{article}
\def\be{\begin{equation}}
\def\ee{\end{equation}}
\def\bea{\begin{eqnarray}}
\def\eea{\end{eqnarray}}
\def\pd{\partial}

\input amssym.def
\input amssym.tex
\begin{document}
\medskip
\makeatletter
\def\eqnarray{\stepcounter{equation}\let\@currentlabel=\theequation
\global\@eqnswtrue
\global\@eqcnt\z@\tabskip\@centering\let\\=\@eqncr
$$\halign to \displaywidth\bgroup\@eqnsel\hskip\@centering
  $\displaystyle\tabskip\z@{##}$&\global\@eqcnt\@ne
  \hfil$\displaystyle{\hbox{}##\hbox{}}$\hfil
  &\global\@eqcnt\tw@ $\displaystyle\tabskip\z@
  {##}$\hfil\tabskip\@centering&\llap{##}\tabskip\z@\cr}
\@addtoreset{equation}{section}
  \def\theequation{\thesection.\arabic{equation}}
\makeatother
\begin{titlepage}

\title{Equations with an infinite number of explicit Conservation Laws}

\author{D.B. Fairlie\\
{\it Department of Mathematical Sciences}\\
{\it University of Durham, Durham DH1 3LE}}
\maketitle

\begin{abstract}
A large class of first order partial nonlinear differential equations in two independent variables which possess an infinite set of polynomial conservation laws 
derived from an explicit generating function is constructed. The
conserved charge densities are all homogeneous polynomials in the unknown functions which satisfy the differential equations in question.
 The simplest member of the
class of equations is related to the Born-Infeld equation in two dimensions. It
is observed that some members of this class possess identical charge densities. This enables the construction of a set of multivariable equations with an
infinite number of conservation laws. 
\end{abstract}
\end{titlepage}

\section{Introduction and examples.}
The purpose of this article is to demonstrate the existence of a class of 
nonlinear first order differential equations in 2 independent variables, but an arbitrary number of dependent ones which admit an infinite number of independent polynomial conservation laws. There are general results which guarantee the existence of an infinite set of conservation laws for evolution equations in 
two dimensions \cite{min}, but there are comparatively few recorded cases where the conserved densities may be written as explicit  polynomials in the dependent variables. One  purpose of this article is to demonstrate a class of equations
in which these conservation laws may be derived from the use of a generating function, another is to exploit these results to exhibit an example with polynomial conservation laws in more than two independent variables.

The equations considered are in fact particular examples of the equations of hydrodynamic type studied by the Russian School of  Novikov and Dubrovin  \cite{nov}\cite {tsar}\cite {dub}, but the explicit nature of the construction
of the conservation laws presented here may be useful. 
For a flavour of the results, consider first the following examples for 4 unknown functions $u_i, \ i=1\dots 4$ which were discovered by a combination of guesswork and computer verification. These results led to a generalisation which will be described  in two stages, first using symmetric polynomials, then 
without symmetry. 
General proofs  will be given in the final two sections.
\subsection{Example 1}
Consider the following first order differential equations
\bea\label{ex1}
\frac{\pd u_1}{\pd t}&=&(u_4+u_2+u_3)\frac{\pd u_1}{\pd x}\nonumber\\
\frac{\pd u_2}{\pd t}&=&(u_1+u_4+u_3)\frac{\pd u_2}{\pd x}\\
\frac{\pd u_3}{\pd t}&=&(u_1+u_2+u_4)\frac{\pd u_3}{\pd x}\nonumber\\
\frac{\pd u_4}{\pd t}&=&(u_1+u_2+u_3)\frac{\pd u_4}{\pd x}.\nonumber
\eea
Construct the matrices $M[n,p,q,r]$ as follows:
\be M[n,p,q,r] =\left|\begin{array}{cccc}
u_1^n&u_2^n&u_3^n&u_4^n\\
u_1^p&u_2^p&u_3^p&u_4^p\\
u_1^q&u_2^q&u_3^q&u_4^q\\
u_1^r&u_2^r&u_3^r&u_4^r\end{array}\right|\label{matrices}
\ee
and define $\Delta{\buildrel \rm def \over =}\det M[3,2,1,0]$.

Then the conservation laws for (\ref{ex1})  are simply
\be
\frac{\pd}{\pd t}\left(\frac{\det M[n,2,1,0]}{\Delta}\right)=\frac{\pd }{\pd x}\left(\frac{\det M[n,3,1,0]}{\Delta}\right)\ \ \forall\ {\rm positive}\ n >3.\label{cons1}
\ee
The conserved densities are clearly linearly independent.
\subsection{Example 2.}
Consider instead the following first order differential equations given by
\be\label{ex2}
\frac{\pd u_1}{\pd t}=(u_4u_2+u_2u_3+u_3u_4)\frac{\pd u_1}{\pd x};
\ee
together with its cyclic replacements.
Then the conservation laws for this system of equations are simply
\be
\frac{\pd}{\pd t}\left(\frac{\det M[n,2,1,0]}{\Delta}\right)=\frac{\pd }{\pd x}\left(\frac{\det M[n,3,2,0]}{\Delta}\right).
\label{cons2}
\ee

\subsection{Example 3.}
Change once again the  first order differential equations as follows;
\be\label{ex3}
\frac{\pd u_1}{\pd t}=u_2u_3u_4\frac{\pd u_1}{\pd x};\ \ {\rm together\ with \ cyclic\ replacements}.
\ee
Now the conservation laws for (\ref{ex3}) become
\be
\frac{\pd}{\pd t}\left(\frac{\det M[n,2,1,0]}{\Delta}\right)=\frac{\pd }{\pd x}\left(\frac{\det M[n,3,2,1]}{\Delta}\right).
\ee

The general pattern becomes clear. The conserved charge density is a symmetric polynomial. It is the same one for each example. This leads to the evident  
consequence that a correponding result holds good for a linear combination of equations (\ref{ex1}), (\ref{ex2}) and (\ref{ex3}). Furthermore, one may take equation
(\ref{ex1}) with independent variables $x,\ t$ and add to the right hand side that of 
(\ref{ex2}) with independent variables $y,\ t$ and (\ref{ex3}) with independent variables $z,\ t$ and thus construct an example of a set of partial differential equations in 4 dimensions, with independent variables $ x,\ y,\,\ z,\ t$
which admits an infinite number of conservation laws. This construction clearly generalises to any number of variables. This point will be elaborated later.
\subsection{Implicit Solutions.}
Example (\ref{ex1}) can be integrated implicitly in the case of two dependent variables.
In the case of three functions $u_1,\ u_2,\ u_3,$ the equations (\ref{ex2})
can be reduced by the hodograph method of interchanging the r\^oles of the independent variables $x,\ t$ and  the dependent variables $u_1,\ u_2$ to give
\bea\label{3dhod}
\frac{\pd x}{\pd u_2}&=&-u_2u_3\frac{\pd t}{\pd u_2}\nonumber\\
\frac{\pd x}{\pd u_1}&=&-u_3u_1\frac{\pd t}{\pd u_1}\\
\frac{\pd u_3}{\pd u_1}\frac{\pd t}{\pd u_2}&=&-\frac{u_1(u_2-u_3)}{u_2(u_3-u_1)}\frac{\pd u_3}{\pd u_2}\frac{\pd t}{\pd u_1}.\nonumber
\eea
Introduce three arbitrary functions of a single variable, $f(u_1),\ g(u_2),$
and $F(z)$ where
\be
 z=\int^{u_1}u'f(u')du'+\int^{u_2}v'g(v')dv'.
\label{def}
\ee
Then the solution of (\ref{3dhod}) is given explicitly by quadratures as
\bea
t&=&\int^{u_1}f(u')du'+\int^{u_2}g(v')dv'-\int^{\int^{u_1}u'f(u')du'+\int^{u_2}v'g(v')dv'}
F(z)^{-1}dz\nonumber\\
x&=&\int^{u_1}u'^2f(u')du'+\int^{u_2}v'^2g(v')dv'-\int^{\int^{u_1}u'f(u')du'+\int^{u_2}v'g(v')dv'}
F(z)dz\nonumber\\
u_3&=&F(\int^{u_1}u'f(u')du'+\int^{u_2}v'g(v')dv').\label{sol}
\eea
\section{Application.}
 In the case of two 
dependent variables,
$u_1,\ u_2$  equations of the form (\ref{ex1}) arose in a study of the Born-Infeld equation
with J.Mulvey\cite{joe}. The argument may be reversed as follows; Suppose
a second order Lorentz invariant equation (regarding $t,\ x$ as light-cone 
co-ordinates) is sought by setting 
\bea\label{born}
u_1&=&\frac{f(\frac{\pd\phi }{\pd t}\frac{\pd\phi }{\pd x})}{(\frac{\pd\phi }{\pd x})^2}\nonumber\\
u_2&=&\frac{g(\frac{\pd\phi }{\pd t}\frac{\pd\phi }{\pd x})}{(\frac{\pd\phi }{\pd x})^2}.
\eea
Then the resulting equations to determine $f,\ g$ 
are so restrictive that
essentially the only solution reproduces the Born-Infeld equation, which
 in 1+1 dimensions is given in terms of the parameter $\lambda$ by
\be
(\frac{\pd\phi}{\pd x})^2\frac{\pd^2\phi}{\pd t^2}+(\frac{\pd\phi}{\pd
t})^2\frac{\pd^2\phi}{\pd x^2}- (\lambda+2\frac{\pd\phi}{\pd
x}\frac{\pd\phi}{\pd t})\frac{\pd^2\phi}{\pd x\pd t}=0.
\label{1}
\ee
and $u_1,\  u_2$ are roots of the characteristic equation
\be
(\frac{\pd\phi}{\pd x})^2u^2- (\lambda+2\frac{\pd\phi}{\pd x}\frac{\pd\phi}{\pd t})u+(\frac{\pd\phi}{\pd t})^2=0.
\ee
\section{First Generalisation}

Let ${ S}_n$ be the $n$th elementary symmetric polynomial  of degree $n$ (i.e. the one in which no variable appears more than linearly in any term) 
in $N$ variables $u_k,\ k=1\dots N,\ n\leq N$ and $S^{\hat j}_n$  be the $n$th elementary symmetric polynomial  of degree $n$ 
in $N-1$ variables $u_k,\ k=1\dots,\hat j,\dots\ N,$ where the superscript $\hat j$
denotes that the variable $u_j$ is omitted.
Postulate the existence of the differential equations
\be
 \frac{\pd u_j}{\pd t}=S^{\hat j}_n\frac{\pd u_j}{\pd x};\ \ \ j=1,\dots,N.
\label{post}
\ee
Introduce the functions $\phi_k,\ k=0\dots N$ given by
\[\displaystyle{\phi_0=\frac{1}{\prod_{i=1}^{i=N}(\lambda-u_i)}}\] 
and the recurrence relations
\be
\phi_k+\lambda\phi_{k-1}=\frac{{S}_k}{\prod_{i=1}^{i=N}(\lambda-u_i)},
\label{eq1}
\ee
with $\phi_N=1$. Then the result is that in consequence of the equations (\ref{post}) the following relation holds
\be
\frac{\pd \phi_0}{\pd t}=\frac{\pd \phi_n}{\pd x}.
\label{generate}
\ee
Comparison of the coefficients of inverse powers of $\lambda$ on both sides of a formal expansion of this equation gives the various conservation laws, since the quantities $\phi_0,\ \phi_n$ are generating functions for the conserved densities and currents respectively. These coefficients will be symmetric polynomials in the dependent variables which can in fact be identified with
the evident extension of the ratios of the determinants of Vandermonde type presented in the three examples.  The proof of these results hinges upon the fact that it is often easier to construct a generating function for a set of conservation laws than the conservation laws themselves\cite{chand}. It runs as follows;
\subsection{Proof}
The relations (\ref{eq1}) can be solved in terms of $\phi_0$ to yield
\be
\phi_k=\frac{\displaystyle{\sum_{r=0}^{r=k}(-1)^{k-r}{ S}_r\lambda^{k-r}}}
{\prod_{i=1}^{i=N}(\lambda-u_i)},
\label{eq2}
\ee
with ${ S}_0=1$. Then using the facts that
\bea\label{fact2}
 \frac{\pd { S}_n}{\pd u_j}&=&S^{\hat j}_{n-1};\label{fact1}\\
u_jS^{\hat j}_{n-1}&=&{ S}_n-S^{\hat j}_n,
\eea
which are easily established, the following relations may be deduced;
\bea
\frac{\pd \phi_n}{\pd x}&=&\sum_j\phi_0(\frac{\pd }{\pd u_j}{\sum_{r=0}^{r=n}(-1)^{n-r}{ S}_r\lambda^{n-r}}+\frac{\displaystyle{\sum_{r=0}^{r=k}(-1)^{k-r}{ S}_r\lambda^{k-r}}}{\lambda-u_j})\frac{\pd u_j}{\pd x}\nonumber\\
&=&\sum_j\phi_0\frac{S^{\hat j}_n}{\lambda-u_j}\frac{\pd u_j}{\pd x}\nonumber\\
&=&\frac{\pd \phi_0}{\pd t}.
\label{two}
\eea
Here relations  (\ref{fact2}), (\ref{fact1}) and equations (\ref{post}) have been used. This is just the result (\ref{generate}) claimed. 

The connection with the specific examples written in determinantal form requires some further explanation.
Consider the quantity $\displaystyle{\frac{\det M[n,2,1,0]}{\Delta}}$.
This is a symmetric polynomial of degree $n-3$ in the variables $ u_1,\ u_2,\ u_3,\
u_4$ such that the coefficient of every term is unity.
Now consider the expansion in powers of $\lambda^{-1}$ of the expression
\be
\phi_0=\frac{\lambda}{(\lambda-u_1)(\lambda-u_2)(\lambda-u_3)(\lambda-u_4)}.\nonumber
\ee
Then the coefficient of $\lambda^{-n}$ in the power series expansion of this
expression is the quantity  $\displaystyle{\frac{\det  M[n,2,1,0]}{\Delta}}$.
This may be seen in various ways, the most direct of which is the observation that the coefficients in the expansion  of each individual factor are all unity, and in the multiplication of the geometrical series all possible terms arise with unit coefficient. It can also be seen from the definition of the matrices
involved, (\ref{matrices}) with the help of the expansion in partial fractions
\bea\label{pole}
\frac{1}{(\lambda-u_1)(\lambda-u_2)(\lambda-u_3)(\lambda-u_4)}&=&
\frac{1}{(\lambda-u_1)(u_1-u_2)(u_1-u_3)(u_1-u_4)}\nonumber\\+\frac{1}{(\lambda-u_2)(u_2-u_1)(u_2-u_3)(u_2-u_4)}&+&\frac{1}{(\lambda-u_3)(u_3-u_1)(u_3-u_2)(u_3-u_4)}\nonumber\\+\frac{1}{(\lambda-u_4)(u_4-u_1)(u_4-u_2)(u_4-u_3)}&,&
\eea
by considering the expansion of the determinant on the first row, which corresponds to taking the coefficient of $\lambda^{-n}$ on the right hand side of this expression. In a similar manner the identification of  $\displaystyle{\frac{\det  M[n,3,1,0]}{\Delta}}$ with the coefficient of 
 $\lambda^{-n}$ in the expansion of
\[\phi_1=\frac{-\lambda+u_1+u_2+u_3+u_4}{(\lambda-u_1)(\lambda-u_2)(\lambda-u_3)(\lambda-u_4)}\]
may be best accomplished by way of the identity (\ref{pole}) and expansions of the determinant on the first row. The results for the other determinants can
be established likewise.
\subsection{Multivariable case}
Since the same function $\phi_0$ enters into the generating equation (\ref{generate}) for each $n$, these relations may be combined linearly to
produce a multivariable extension; Suppose there are $N$ independent variables
$t,\ x_i,\ i=1\dots N-1$ and the $N$ dependent variables $u_j,\  j=1\dots N$ satisfy the quasi-linear system of equations
\be
 \frac{\pd u_j}{\pd t}=\sum_{n=1}^{N-1}S^{\hat j}_n\frac{\pd u_j}{\pd x_n};\ \ \ j=1,\dots,N.
\label{post1}
\ee
Then this set of equations admits an infinite number of linearly independent polynomial conservation laws, obtained by equating  the coefficients of  inverse powers of $\lambda$ in a formal expansion of the generating function 
\be
\frac{\pd \phi_0}{\pd t}=\sum_{n=1}^{N-1}\frac{\pd \phi_n}{\pd x_n}.
\label{mgenerate}
\ee
It is not clear at this stage whether the existence of these laws is sufficient to guarantee integrability of (\ref{post1}). 
\section{Further asymmetric extension}
The system admits a further generalisation which breaks the symmetry.
Consider the following system of first order differential equations;
\bea\label{ex5}
\frac{\pd u_1}{\pd t}&=&((n_1-1)u_1+n_2u_2+n_3u_3)\frac{\pd u_1}{\pd x}\nonumber\\
\frac{\pd u_2}{\pd t}&=&(n_1u_1+(n_2-1)u_2+n_3u_3)\frac{\pd u_2}{\pd x}\\
\frac{\pd u_3}{\pd t}&=&(n_1u_1+n_2u_2+(n_3-1)u_3)\frac{\pd u_3}{\pd x}.\nonumber
\eea
This system possesses an infinite number of conservation laws each individual
member of which corresponds to equating coefficients of $\lambda^{-k}$ on both sides of  the equation
\be
\frac{\pd}{\pd t}\frac{1}{(\lambda-u_1)^{n_1}(\lambda-u_2)^{n_2}(\lambda-u_3)^{n_3}}=
\frac{\pd}{\pd x}\frac{-\lambda+n_1u_1+n_2u_2+n_3u_3}{(\lambda-u_1)^{n_1}(\lambda-u_2)^{n_2}(\lambda-u_3)^{n_3}}.\label{asym}
\ee
This example may be further generalised by  scaling $u_1,\ u_2,\ u_3$ to
make the `diagonal' entries on the right of equations (\ref{ex5}) arbitrary.
If a solution of these equations of the form $u_i=xf_i(t)$, then these equations assume the typical form of dynamical evolution equations. 
The following general result may be abstracted from (\ref{ex5}) and the previous examples.
Consider the expression
\be
\Phi_M=\frac{\displaystyle{\sum_{j=0}^M\oint\prod_{k=1}^{k=N} (\lambda-zu_k)^{n_k}\frac{dz}{2\pi iz^{j+1}}}}{\displaystyle{\prod_{k=1}^{k=N}(\lambda-u_k)^{n_k}}}.
\label{nef}
\ee
The construction in the numerator is just a  device to take the sum of only  the first
$M+1$ terms in the expansion of ${\displaystyle{\prod_{k=1}^{k=N}(\lambda-u_k)^{n_k}}}$
 in decreasing powers of $\lambda$. It is assumed that if any of the $n_j$ are not positive integers then
the contour integral is taken round a contour including the origin, but excluding any other possible singularities.
Then 
\bea\label{big}
\frac{\pd\Phi_M}{\pd u_i}&=&\frac{\Phi_0}{\lambda^{\sum n_j}(\lambda-u_i)}\left((\lambda-u_i)\frac{\pd}{\pd u_i}(\sum_{j=0}^M\oint\prod_{k=1}^{k=N} (\lambda-zu_k)^{n_k}\frac{dz}{2\pi iz^{j+1}})\right)\nonumber\\
&+&n_i\frac{\Phi_0}{\lambda^{\sum n_j}(\lambda-u_i)}\left(\sum_{j=0}^M\oint\prod_{k=1}^{k=N} (\lambda-zu_k)^{n_k}\frac{dz}{2\pi iz^{j+1}}
\right)\nonumber\\
&=&n_i\frac{\Phi_0}{\lambda^{\sum n_j}(\lambda-u_i)}\left(\sum_{j=0}^M\oint\frac{\lambda (1-z)\prod_{k=1}^{k=N} (\lambda-zu_k)^{n_k}}{{2\pi iz^{j+1}}(\lambda-zu_i)}dz
\right)\\
&=&n_i\frac{\Phi_0}{\lambda^{\sum n_j}(\lambda-u_i)}\oint\frac{\lambda\prod_{k=1}^{k=N} (\lambda-zu_k)^{n_k}}{2\pi i(\lambda-zu_i)}(z^{-(M+1)}-1) dz\nonumber\\
&=&\lambda^{-M}\frac{\pd\Phi_0}{\pd u_i}\times{\rm\  function\ independent\ of}\ \lambda.\nonumber
\eea
Note that the index i is free in (\ref{big}).
In the last two steps first the geometric
series is summed, then the contour integral performed yielding a function
which after removing a factor $\displaystyle{\lambda^{-M}}$ is independent of $\lambda$. This function is, as can be seen from (\ref{big}), equivalent to setting $\lambda=u_i$ in  the expression
$\displaystyle{\sum_{j=0}^M\oint\prod_{k}^{k=N} (1-z\frac{u_k}{\lambda})^{n_k}\frac{dz}{2\pi iz^{j+1}}}$.

These considerations give the final generalisation; The function
$\Phi_0$ serves as a generating function for the conservation laws for a very wide class of nonlinear differential equations as a consequence of the result

\be
\frac{
\pd\Phi_0}{\pd t}=
\lambda^M\frac{\pd\Phi_M}{\pd x}
\label{fin1}
\ee
where the dependent functions $u_i$ satisfy the equations
\be
\frac{\pd u_i}{\pd t}=
\frac{\pd u_i}{\pd x}\left.(\sum_{j=0}^M\oint\prod_{k\neq i}^{k=N} (1-z\frac{u_k}{\lambda})^{n_k}\frac{dz}{2\pi iz^{j+1}})\right|_{\lambda=u_i}.
\label{fin2}
\ee
The individual conservation laws are obtained by equating powers of $\displaystyle{\frac{1}{\lambda}}$ on both sides of (\ref{fin1}). In the case 
where all $n_i$ are positive integers, this final generalisation is equivalent to the symmetric case when subsets of the variables $u_i$ are identified. 
If a solution with power law $x$ dependence of the equations of the type
(\ref{fin2}) is sought, then these equations become nonlinear evolution
 equations of the type studied in dynamical systems theory. For example, if $u_j(x,t)=xv_j(t)$ then the $x$ dependence of the equations (\ref{ex5}) drops out. 
\section{Coda}
A recent attempt to prove the conservation laws for example 1 directly
led to the further generalisation;
\[
\frac{\partial}{\partial t}\det\left|\begin{array}{cccc}
f(u_1)&g(u_2)&h(u_3)&k(u_3)\\
u_1^2&u_2^2&u_3^2&u_4^2\\u_1&u_2&u_3&u_4\\
\frac{1}{\Delta}&\frac{1}{\Delta}&\frac{1}{\Delta}&\frac{1}{\Delta}\end{array}\right|=\frac{\partial}{\partial x}\det\left|\begin{array}{cccc}
f(u_1)&g(u_2)&h(u_3)&k(u_3)\\
u_1^3&u_2^3&u_3^3&u_4^3\\u_1&u_2&u_3&u_4\\
\frac{1}{\Delta}&\frac{1}{\Delta}&\frac{1}{\Delta}&\frac{1}{\Delta}\end{array}\right|
\]
This equation is a conservation law for equations (\ref{ex1}) for any arbitrary differentiable functions $f,\ g,\ h,\ k$. This result may be proved by straightforward differentiation after expansion on the first row of the determinants involved. Clearly this result will generalise.

\end{document}